\documentclass{iaus}
\usepackage{graphicx}

\title[Super Massive Star Clusters] %% give here short title %%
{Super Massive Star Clusters: From Superwinds to a Cooling 
               Catastrophe and the Re-processing of the Injected Gas }

\author[Silich et al.]   %% give here short author list %%
{Silich S.$^1$  \break 
G. Tenorio-Tagle$^1$ C. Mu\~noz-Tu\~n\'on $^2$ J. Palou\v{s},$^{3}$}

\affiliation{$^1$ Instituto Nacional de Astrof\'\i sica Optica y
Electr\'onica, AP 51, 72000 Puebla, M\'exico \break email: silich@inaoep.mx
\\[\affilskip]
$^2$ Instituto de Astrof\'{\i}sica de Canarias, E 38200 La
Laguna, Tenerife, Spain \break cmt@ll.iac.es
\\[\affilskip]
Astronomical Institute, Academy of Sciences of the Czech
Republic, Bo\v{c}n\'\i\ II 1401, 141 31 Prague, Czech Republic 
\break email: palous@ig.cas.cz}

\pubyear{2006}
\volume{237}  %% insert here IAU Symposium No.
\pagerange{0--4}
\date{?? and in revised form ??}
\jname{Triggered Star Formation in a Turbulent ISM}
\editors{B.G. Elmegreen \& J.  Palou\v{s}, eds.}
\begin{document}

\maketitle

\begin{abstract}
Different hydrodynamic regimes for the gaseous outflows generated by multiple 
supernovae explosions and stellar winds occurring within compact and massive 
star clusters are discussed. It is shown that there exists the threshold
energy that separates clusters whose outflows evolve in the quasi-adiabatic 
or radiative regime from those within which catastrophic cooling and a 
positive feedback star-forming mode sets in. The role 
of the surrounding ISM and the observational appearance of the star cluster 
winds evolving in different hydrodynamic regimes are also discussed. 
\keywords{galaxies: star clusters, galaxies: starburst, ISM: jets and 
outflows}
\end{abstract}

\firstsection % if your document starts with a section,
              % remove some space above using this command.
\section{Introduction}

In many starburst galaxies, in interacting and merging galaxies, a 
substantial fraction of star formation is concentrated in a number of 
compact, young and massive stellar clusters (SSCs) which may represent 
the earliest stages of globular clusters evolution. In the extreme scenario 
SSCs represent the dominant mode of star formation in these galaxies
(see, for example, \cite{McCrady03}; \cite{smith06} and references therein). 
Powerful gaseous outflows associated with such clusters are
now believed to be one of the major agents leading to a large-scale structuring
of the ISM in the host galaxies and to the dispersal of heavy elements into the
ISM and the IGM.

Analysis of the SSC's outflows led us to realize that radiative cooling may 
crucially affect the hydrodynamics of the star cluster winds and that the
superwind concept proposed by \cite{CC85} required a substantial modification
in the case of very massive and compact star clusters. We demonstrated that
there exists the threshold line in the mechanical energy input rate vs the 
cluster size parameter space. This line separates clusters whose outflows 
evolve in the quasi-adiabatic or radiative regime from those in which   
catastrophic cooling sets in inside the cluster. In the catastrophic cooling 
regime (above the threshold line) at least some fraction of the matter
reinserted via strong stellar winds and supernovae 
remains bound within  the cluster and is finally re-processed into new
generations of stars (see \cite{GTT2005}; \cite{W2006a}). Here we review the 
subject and  discuss also how the high 
pressure in the surrounding medium may prohibit the development of high 
velocity star cluster winds turning them into low velocity,  
subsonic outflows. Finally, we make some predictions regarding the 
observational manifestations of the star cluster outflows evolving in the 
different hydrodynamic regimes.

\section{The threshold mechanical luminosity}\label{sec:1}

In the stationary regime the injection of matter by supernovae and
stellar winds, ${\dot M}_{SC}$, is balanced by the mass outflow 
driven by the large cental overpressure which results from the
efficient thermalization of the kinetic energy, $ L_{SC}$, deposited by SNe
and stellar winds:
%---------------------------------------------------------------
\begin{equation}
\label{eq.1}
{\dot M}_{SC} = 4 \pi R^2_{SC} \rho_{SC} a_{SC} ,  
\end{equation} 
%---------------------------------------------------------------
where $R_{SC}$ is the radius of the cluster, $\rho_{SC}$ is the
density of the out-flowing gas at the star cluster surface, and
$a_{SC} \approx V_{A\infty} / 2$ is the speed of sound at the star
cluster edge. $V_{A\infty} = (2 L_{SC} / {\dot M}_{SC})^{1/2}$ is
the adiabatic wind terminal speed. Equation \ref{eq.1} indicates
that the density of the plasma inside more massive clusters is
larger,  if other
parameters ($R_{SC}, V_{A\infty}$) do not change.
This implies that the impact of radiative cooling on the star cluster winds 
becomes progressively more important for more massive clusters.

\cite{silich04} and \cite{GTT2005} have demonstrated that for larger or more 
massive clusters, the larger the leakage of thermal energy. This leads  
first to a sharp drop in the wind temperature at some distance from the star 
cluster surface. When the mechanical energy input rate, $L_{SC}$, exceeds the 
critical value, the stationary wind solution is inhibited. The critical power 
is defined  by the condition that the central pressure reaches the maximum
value, allowed by the radiative cooling.
The critical energy crucially depends on the thermalization or 
heating efficiency, $e_t$. This parameter 
characterizes how efficient the transformation of the mechanical energy
supplied by supernovae and stellar winds,  into  
thermal energy is (Stevens \& Hartwell, 2003; 
Melioli \& Del Pino, 2004). 

The threshold line calculated under assumption that $V_{A\infty} = 1500$
km s$^{-1}$ for two cases,  $e_t = 1$ and $e_t = 0.1$, together with
several massive SSCs, is presented in figure 1a.
%-------------------------------------------------------------------
\begin{figure}
\vspace{7.5cm}
\includegraphics{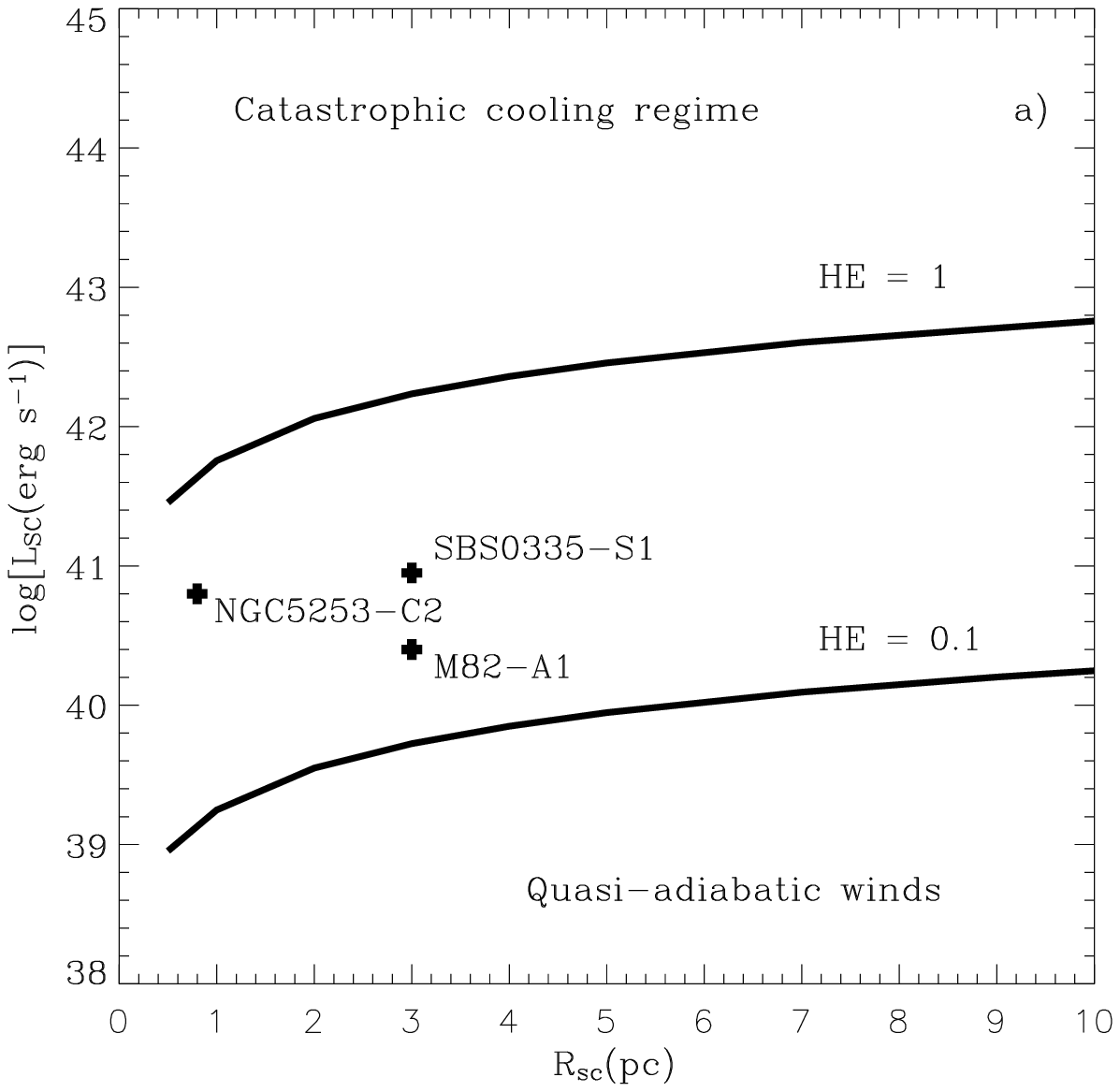}
\includegraphics{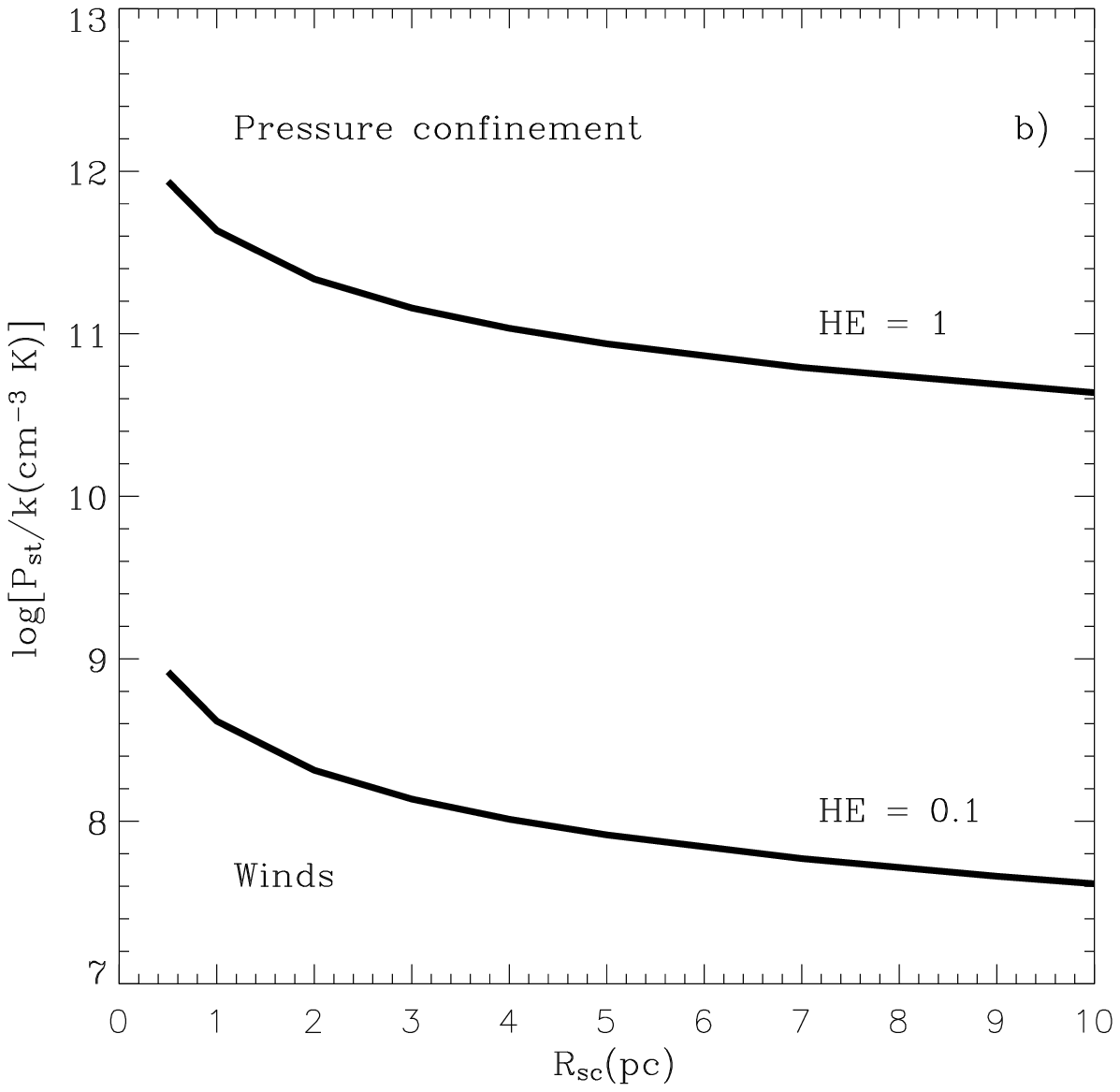}
\caption{The threshold line. Panel a) presents the
critical energy input rates for clusters with different radii. Below the
threshold line the stagnation point accommodates at the star cluster
center and all mass supplied by massive stars  conforms a star cluster wind. 
Above the threshold
line the stagnation point moves to some distance away from the star cluster
center (see \cite{W2006a}) and only a fraction of the deposited mass and
energy leave the cluster. The calculations were done for two values of the 
heating efficiency, $e_t = 1$ (the upper line), and  $e_t = 0.1$
(the lower line). Panel b) presents the pressure at the
stagnation point for clusters with critical energy input rates,
$P_{thresh}$. The upper line was calculated for $e_t = 1$
and the lower one is for $e_t = 0.1$. Using these two diagrams one can 
obtain the pressure at the stagnation point for any cluster above the 
threshold line (see equation \ref{eq.3}). The locations of several 
massive SSCs with respect to the threshold lines are indicated in panel a)
by the cross symbols.}  
\end{figure}
%-------------------------------------------------------------------

\section{The impact of the external pressure}\label{sec:2}

Radiative cooling puts important restrictions on the plasma parameters inside 
the cluster. In particular, the pressure at the stagnation point ($R_{st}$; 
the point where the expansion velocity is equal to zero; below the threshold 
line $R_{st} = 0$), which is the largest across the cluster,
%---------------------------------------------------------------
\begin{equation}
      \label{eq.2}
P_{st} = k T_{st} q^{1/2}_m \left[\frac{V^2_{A\infty}/2 - a^2_{st} / 
         (\gamma - 1)}{\Lambda(T_{st},Z)}\right]^{1/2} , 
\end{equation}
%---------------------------------------------------------------
is restricted by the shape of the cooling function, $\Lambda(T,Z)$, and  the 
cluster's  parameters, $q_m = 3 {\dot M}_{SC} / 4 \pi R^3_{SC}$ and  
$V_{A\infty}$ (\cite{silich04}). Figure 1b displays
$P_{st}$ for clusters with critical mechanical luminosities as presented
in figure 1a. Figures 1a and 1b allow one to calculate the pressure at the
stagnation point for any cluster above the threshold line without knowing
the location of the stagnation radius:
%---------------------------------------------------------------
\begin{equation}
      \label{eq.3}
P_{st} = P_{thresh} (L_{SC} / L_{thresh})^{1/2} ,
\end{equation}
%---------------------------------------------------------------
where $L_{thresh}$ is presented in figure 1a, and $P_{thresh}$  
is the pressure at the stagnation point along the threshold line (figure 1b).
If the pressure in the surrounding interstellar medium exceeds that at the
stagnation point, $P_{ISM} \ge P_{st}$, the cluster is not able to blow 
away the inserted matter and drive a high velocity outflow. Such clusters 
instead of driving a high velocity wind remain buried by the high 
pressure surrounding medium and effectively reprocess all infalling and 
reinserted mass into stars. Thus the high pressure 
ISM may prevent negative feedback and lead to a  high
star formation efficiency.

\section{Observational appearance of SSC's winds}

The efficient thermalization of the kinetic energy deposited by stellar
winds and supernovae explosions results in a high temperature of the plasma 
within the cluster. 
The distributions of density and temperature inside the high velocity
outflow defines then the observational manifestations of the star 
cluster wind in the different energy bands.

Well below the threshold line, in the quasi-adiabatic regime, the
temperature rapidly reaches its asymptotic trend, $T \sim r^{-4/3}$,
whereas the density drops as $\rho \sim r^{-2}$. Thus well below the
threshold line star cluster winds should be associated with the diffuse 
X-ray sources with the hard component concentrated to the star cluster
volume (see, for example, \cite{CR2000}; \cite{silich05}). In the
visible line regime such winds are hardly to be detected due
to the fast decrease of density and the negligible emission measure
at the distance where the temperature reaches $\sim 10^4$K when the 
outflowing plasma begins to recombine and produce an emission line spectrum.

However the temperature structure of the outflowing matter changes
drastically for clusters approaching or located above the threshold 
line.  For such clusters the temperature falls down and rapidly reaches 
$10^4$ K at much smaller distances from the star cluster surface.
This warm gas, photoionized by the star cluster Lyman continuum, 
moves with velocity around 1000 km s$^{-1}$ and should be detected as
a low intensity broad line emission.   

When $L_{SC}$ exceeds the threshold value, the catastrophic cooling sets in 
first at the center. The cooling front and the stagnation point then move from 
the star cluster center outwards (see \cite{W2006a}). The temperature of the 
initially thermalized material then rapidly drops from $\sim 10^7$~K to 
approximately $10^4$ K where it is balanced by the ionizing radiation from  
massive stars. The density of the photoionized material grows larger  until 
the gravitational instability sets in and the accumulated gas begins to form 
new stars (\cite{GTT2005}). The emission line spectra from such clusters 
should present the central narrow peak associated with the dense gas
inside the star forming region which is located inside the stagnation
radius, $R_{st}$, and the lower intensity broad component associated
with the warm, photoionized, fast outflow. The two regions are separated
by a shell of the hot, thermalized plasma which should be detected as 
1.0 - 4.0 keV X-ray emission.    

\begin{acknowledgments}
We would like to acknowledge the support of this study given by
CONACYT - M\'exico, research grant 
47534-F and AYA2004-08260-CO3-O1 from the Spanish Consejo Superior de
Investigaciones Cient\'\i{}ficas.
\end{acknowledgments}

\begin{discussion}

\discuss{S. Dib}{Can you please comment on the role of metallicity on your 
models - for example if the metallicity is subsolar?}

\discuss{S. Silich}{The cooling rate is larger in a plasma with higher
metal abundance. Thus the enhanced metallicity will shift the threshold  
energy towards lower values and vice versa. Note however that we are dealing 
with matter returned to the ISM through winds and supernovae and thus we 
expect the metallicities to be large.}

\end{discussion}

\end{document}